\documentclass[final,1p,times]{elsarticle}
\usepackage{graphicx}
\usepackage{picinpar}
\usepackage{amsmath}
\usepackage{url}
\usepackage{flushend}
\usepackage[latin1]{inputenc}
\usepackage{colortbl}
\usepackage{soul}
\usepackage{multirow}
\usepackage{pifont}
\usepackage{color}
\usepackage{alltt}
\usepackage[hidelinks]{hyperref}
\usepackage{enumerate}
\usepackage{enumitem}
\usepackage{siunitx}
\usepackage{breakurl}
\usepackage{epstopdf}
\usepackage{pbox}
\usepackage{booktabs}
\usepackage{amsmath,amssymb}
\usepackage{stfloats}
\usepackage{tabularx,booktabs} 
\usepackage{array}
\usepackage{booktabs}
\usepackage{array}
\usepackage{pdflscape}    
\newcolumntype{L}[1]{>{\raggedright\arraybackslash}p{#1}}
\newcolumntype{M}[1]{>{\centering\arraybackslash}p{#1}}
\newcolumntype{L}[1]{>{\raggedright\arraybackslash}p{#1}}

\usepackage{rotating}      
\usepackage{booktabs}
\usepackage{array}
\newcolumntype{L}[1]{>{\raggedright\arraybackslash}p{#1}}
\newcolumntype{M}[1]{>{\centering\arraybackslash}p{#1}}

\usepackage{amssymb}

\usepackage{amsmath}
\usepackage{setspace}
\journal{xxx}

\begin{document}

\begin{frontmatter}



\title{Deep Reinforcement Learning in Applied Control: Challenges, Analysis, and Insights}

\author{Klinsmann Agyei}
\affiliation{organization={University of Hertfordshire},
            addressline={School of Physics, Engineering and Computer Science}, 
            city={Hatfield},
            postcode={AL10 9AB}, 
            state={Hertfordshire},
            country={United Kingdom}}
\ead{k.agyei@herts.ac.uk}

\author{Pouria Sarhadi}

\ead{p.sarhadi@herts.ac.uk}

\author{Daniel Polani}

\ead{d.polani@herts.ac.uk}

\begin{abstract}
Over the past decade, remarkable progress has been made in adopting deep neural networks to enhance the performance of conventional reinforcement learning. A notable milestone was the development of Deep Q-Networks (DQN), which achieved human-level performance across a range of Atari games, demonstrating the potential of deep learning to stabilise and scale reinforcement learning. Subsequently, extensions to continuous control algorithms paved the way for a new paradigm in control, one that has attracted broader attention than any classical control approach in recent literature. These developments also demonstrated strong potential for advancing data-driven, model-free algorithms for control and for achieving higher levels of autonomy. However, the application of these methods has remained largely confined to simulated and gaming environments, with ongoing efforts to extend them to real-world applications. Before such deployment can be realised, a solid and quantitative understanding of their performance on applied control problems is necessary. This paper conducts a comparative analysis of these approaches on four diverse benchmark problems with implementation results. This analysis offers a scrutinising and systematic evaluation to shed light on the real-world capabilities and limitations of deep reinforcement learning methods in applied control settings.
\end{abstract}


\end{frontmatter}



\section{Introduction}

Automatic control theory has undergone more than a century of development in modern history, with a myriad of real-life applications in various industries that extend into our daily lives \cite{maxwell1868governors,minorsky1922directional,bennett2002controlhistory}. Most controllers have traditionally been designed based on models or detailed knowledge of the system. Despite many ups and downs throughout its history, the successful deployment of model-based controllers remains unquestionable. Nevertheless, a renewed interest has emerged in data-driven, model-free controllers, motivated by ambitious new goals \cite{saviolo2023datadrivenmodelfree,khaki2023datadriven,tang2022datadriven}. These objectives are not aimed at surpassing the already resolved problems of model-based controllers, but rather at achieving broader ambitions:
\begin{enumerate}[topsep=0pt]
    \item Enhancing the generalisability of controllers beyond existing robust and adaptive methods, which are applied to specific model structures and their associated uncertainties;
    \item Addressing problems that cannot be easily modelled by conventional methods, particularly in the development of autonomous and intelligent control systems where decision-making is central (i.e., increasing the autonomy levels of algorithms \cite{antsaklis2020autonomylevels,vagia2016autonomylevels}).
\end{enumerate}

Such challenges include scenarios where control objectives are either poorly defined or cannot be properly specified, necessitating a shift from 'how' to 'what' to control \cite{chen2024goalwhathow}, or even towards intrinsic motivation to intelligently determine the most appropriate goal \cite{tiomkinpolani2024intrinsic}, potentially realising human-level decision-making and control. Recent advances in Artificial Intelligence (AI), Machine Learning (ML), and their subset, Deep Reinforcement Learning (DRL), have demonstrated significant progress, sometimes achieving superhuman performance, particularly in gaming and simulation environments. These advancements have been more prominent in areas such as text, voice, and image processing, particularly demonstrated with the emergence of Large Language Models (LLMs). However, this progress has not yet been convincingly replicated in the control of dynamical systems, although some promising results have been reported. In the field of DRL, several solutions, developed through a paradigm distinct from classical control theory, have attracted considerable attention in the literature, with encouraging results in various control and autonomy applications \cite{kendall2019learningdrivingaday,eschmann2024learningtoflyinseconds,Hadi2024AUVFormation}. The development of DRL algorithms has seen remarkable advancement, beginning with DQN's breakthrough performance on Atari games by DeepMind \cite{mnih2013playing}. For continuous control problems, DDPG \cite{lillicrap2015continuous} pioneered an actor-critic approach with deterministic policy gradients, whilst TD3 \cite{he2020wd3} enhanced stability by addressing overestimation bias. PPO \cite{schulman2017proximal} introduced a stochastic policy approach, and TD-MPC2 \cite{hansen2023td} merged model predictive control principles with deep learning for improved long-term performance. TD-MPC2 \cite{hansen2023td} extends TD-MPC by integrating decoder-free latent-space with a unified architecture that demonstrates scalability and generalization capabilities across diverse control tasks.
 
Nonetheless, these approaches' success has not yet translated into widespread industrial adoption \cite{tang2025DRLsuccesssurvey}, unlike the enduring fame of the Proportional-Integral-Derivative (PID) controller or, subsequently, the Model Predictive Control (MPC) approach \cite{hagglund2024PIDMPC}. Regardless of the complexities of tuning and implementation, a key challenge is the lack of a clear understanding of the capabilities and limitations of DRL approaches when applied to the operational demands of control, especially in a quantified manner. Some efforts have been made to address this gap through qualitative surveys, such as in \cite{bucsoniu2018RLcomparison} or in specific process control applications focused mainly on tracking performance. However, a systematic and rigorous quantified analysis, as presented in this paper, has not yet been carried out.  There have been a number of recent efforts, primarily in the process control community, aimed at quantitatively assessing the performance of these approaches \cite{lawrence2022deep,dutta2023reinforcement,joshi2023tasac,oh2024quantitative}. Nevertheless, as thoroughly reviewed in Section III, these studies remain limited in scope, and this paper aims to fill that gap by providing a more systematic and comprehensive evaluation.

This paper addresses a critical research question: How do leading model-free DRL algorithms perform when confronted with real-world control challenges, including actuator constraints, time delays, parameter uncertainties, and external disturbances? We evaluate four well-established DRL algorithms DDPG, TD3, PPO, and TD-MPC2, and a model-based LQR controller, all trained with the same reward(cost) function, on four distinct benchmark problems selected for their representative challenges: a non-minimum phase system, the two-mass spring system, a nonlinear autonomous underwater vehicle model, and real-time implementation on a quadrotor drone.

Our contribution includes standardised evaluation framework building on the research in \cite{sarhadi2025standard} using control-specific metrics rather than generic reinforcement learning rewards. Unlike previous studies that typically evaluate algorithms under idealised conditions with perfect models and no practical constraints, our approach incorporates realistic implementation challenges including input time delays, measurement noise, and external disturbances. By applying same testing protocols across all algorithms under these practical conditions, we provide quantitative cross-algorithm comparisons under conditions directly relevant to industrial applications. This systematic analysis aims to bridge the gap between reinforcement learning research and practical control engineering implementation.

\section{Control Algorithms and state-of-the-art in this field}
In this section, we present the control methods evaluated in our benchmarking study, including a classical optimal control baseline (LQI) and four deep reinforcement learning approaches (DDPG, TD3, PPO, and TD-MPC2). We selected these specific algorithms because they represent the spectrum of current state-of-the-art approaches to continuous control: from purely model-free methods (both deterministic and stochastic) to hybrid model-based approaches, allowing us to evaluate which fundamental algorithmic paradigms perform best across our challenging benchmark problems that feature different control characteristics, from fundamental limitations to nonlinearities and complex dynamics. This selection also spans both deterministic and stochastic policy representations, on-policy and off-policy learning, and different approaches to handling stability-performance tradeoffs, providing valuable insights for practical control applications. While a comprehensive explanation of each algorithm is beyond the scope of this work, we focus on essential aspects relevant to our comparative analysis and avoid repeating well-established material; readers are referred to the original sources for full algorithmic details, and the provided codes.

\subsection{Linear Quadratic Integral Control (LQI)}
The LQI controller serves as the classical baseline, offering a well-established optimal model-based solution. LQI augments the original plant system $\dot{x}_p = A_p x_p + B_p u$ with an integrator to eliminate steady-state tracking errors. The augmented system becomes:
\begin{equation}
\large
    \dot{x} = A x + B u_c
\end{equation}
where the augmented state vector is 
\begin{equation}
\large
x = \begin{bmatrix} x_p^\top & \displaystyle\int e \, \mathrm{d}t \end{bmatrix}^\top
\end{equation}
with $e$ being the tracking error, and:
\begin{equation}
\large
    A = \begin{bmatrix} A_p & 0 \\ -C_p & 0 \end{bmatrix}, \quad B = \begin{bmatrix} B_p \\ 0 \end{bmatrix}
\end{equation}

The control input is computed as:
\begin{equation}
\large
    u_c = -K x
\end{equation}
where $K$ is the optimal feedback gain obtained by solving the Riccati equation that minimises:
\begin{equation}
\large
    J = \int \left(x^\top Q x + u_c^\top R u_c\right) \, \mathrm{d}t
\end{equation}
with $Q \succeq 0$ and $R \succ 0$. The integral action ensures zero steady-state error for step references, making LQI superior to LQR for tracking applications. To enable fair comparison, all DRL algorithms are trained using a reward function defined as the negative of this LQI cost function:
\begin{equation}
\large
    r(x, u_c) = -(x^\top Q x + u_c^\top R u_c)
\end{equation}
This approach ensures that all controllers optimise the same underlying objective, facilitating meaningful performance comparisons across different algorithmic approaches.

\subsection{Deep Deterministic Policy Gradient (DDPG)}
DDPG \cite{lillicrap2015continuous} combines policy gradient techniques with insights from Q-learning to address continuous control problems. The algorithm employs two neural networks: an actor network $\mu_{\phi}(s)$ that maps states to deterministic actions and a critic network $Q_{\theta}(s,a)$ that estimates state-action values.

The actor is updated by applying the chain rule to the expected return:
\begin{equation}
\large
\nabla_{\phi} J \approx \mathbb{E}_{s \sim \mathcal{D}} \left[ \nabla_a Q_{\theta}(s,a)\big|_{a=\mu_{\phi}(s)} \nabla_{\phi}\mu_{\phi}(s) \right]
\end{equation}

The critic is updated using temporal difference learning:
\begin{equation}
\large
L(\theta) = \mathbb{E}_{(s,a,r,s') \sim \mathcal{D}} \left[ \left( r + \gamma Q_{\theta'}(s', \mu_{\phi'}(s')) - Q_{\theta}(s, a) \right)^2 \right]
\end{equation}

where $Q_{\theta'}$ and $\mu_{\phi'}$ represent the target critic and actor networks, respectively. While DDPG has demonstrated impressive performance, it suffers from value overestimation and sensitivity to hyperparameters, which the following algorithm addresses.

\subsection{Twin Delayed Deep Deterministic Policy Gradient (TD3)}
TD3 \cite{fujimoto2018addressing} improves DDPG through three key mechanisms designed to address specific limitations of the original algorithm. First, TD3 maintains multiple critic networks and uses the minimum of their estimates to reduce overestimation bias:
\begin{equation}
\large
y = r + \gamma \min_{i=1,2} Q_{\theta'_i}(s', \tilde{a})
\end{equation}
Using multiple critics provides a balance between computational efficiency and bias reduction, whilst additional critics could further reduce overestimation, the diminishing returns typically do not justify the increased computational cost.

Secondly, the actor network is updated less frequently than the critic networks, allowing the critics to provide more accurate value estimates before policy improvement. This delayed update prevents the policy from exploiting poor value estimates during early training phases.

Finally, target policy smoothing adds noise $\epsilon$ to target actions to prevent exploitation of critic approximation errors:
\begin{equation}
\large
\tilde{a} = \mu_{\phi'}(s') + \mathrm{clip}(\epsilon, -c, c), \quad \epsilon \sim \mathcal{N}(0, \sigma)
\end{equation}
where the noise $\epsilon$ forces the critic to learn values over a neighbourhood of actions rather than sharp peaks, improving robustness. The clipping bounds $c$ ensure the noisy actions remain within feasible ranges.

These improvements lead to more stable learning and better performance over DDPG, particularly for complex control tasks with high-dimensional state and action spaces.

\subsection{Proximal Policy Optimization (PPO)}
PPO \cite{schulman2017proximal} addresses the challenge of maintaining stable policy updates in reinforcement learning through a clipped surrogate objective function. Unlike the deterministic policies of DDPG and TD3, PPO employs a stochastic policy $\pi_{\theta}(a|s)$ that outputs a probability distribution over actions, enabling natural exploration without requiring explicit noise injection.

The core innovation of PPO lies in its clipped objective function that prevents excessively large policy updates:
\begin{equation}
\large
L^{CLIP}(\theta) = \mathbb{E}_t \left[ \min\left( r_t(\theta) \hat{A}_t, \text{clip}(r_t(\theta), 1-\epsilon, 1+\epsilon) \hat{A}_t \right) \right]
\end{equation}
where $r_t(\theta) = \frac{\pi_{\theta}(a_t|s_t)}{\pi_{\theta_{old}}(a_t|s_t)}$ represents the probability ratio between new and old policies, $\hat{A}_t$ is the advantage estimate, and $\epsilon$ is the clipping parameter that constrains policy changes.

The clipping mechanism ensures that the policy ratio remains within the bounds $[1-\epsilon, 1+\epsilon]$, preventing destructively large updates that could destabilise learning. This approach provides a practical balance between sample efficiency and training stability, making PPO particularly robust to hyperparameter choices.

PPO combines this clipped objective with on-policy learning, updating the policy using trajectories collected from the current policy. Whilst this approach may sacrifice some sample efficiency compared to off-policy methods, it offers a better stability and reliability, making it well-suited for control applications where consistent performance is crucial.

\subsection{Temporal Difference Learning for Model Predictive Control 2 (TD-MPC2)}
TD-MPC2 represents a significant advancement in model-based reinforcement learning, combining temporal difference learning with model predictive control in a unified architecture with a better generalisation capabilities.
The algorithm encodes observed states into a latent representation:
\begin{equation}
\large
z_t = e_\phi(s_t)
\end{equation}
It then learns three core components in this latent space:
Dynamics Model: Predicts state transitions when actions are applied:
\begin{equation}
\large
z_{t+1} = f_\phi(z_t, a_t)
\end{equation}

Reward Model: Estimates immediate rewards for state-action pairs:
\begin{equation}
\large
\hat{r}_t = r_\phi(z_t, a_t)
\end{equation}

Value Function: Predicts expected future returns from latent states:
\begin{equation}
\large
V_\psi(z_t) \approx \mathbb{E}\left[\sum_{k=0}^{\infty} \gamma^k r_{t+k} \right]
\end{equation}

For planning, TD-MPC2 uses the Cross-Entropy Method (CEM) to optimise action sequences over a finite horizon $H$:
\begin{equation}
\large
\mathbf{a}_{0:H-1}^* = \underset{\mathbf{a}_{0:H-1}}{\arg\max} \sum_{i=0}^{H-1} \gamma^i r_\phi(z_{t+i}, a_{t+i}) + \gamma^H V_\psi(z_{t+H})
\end{equation}
subject to the learned dynamics.

TD-MPC2's key innovation is its decoder-free architecture operating directly in latent space, enabling good sample efficiency and performance, particularly in systems with complex dynamics where explicit reasoning about future states provides significant advantages.

\subsection{State-of-the-art progress in DRL assessment for control application}
\label{subsec:review}
Several studies have applied reinforcement learning to control problems, but a comprehensive evaluation of model-free DRL algorithms' performance on standardised control benchmarks remains a significant gap in the literature. As summarised in Table \ref{tab:rl_process_control}, existing work has primarily focused on specific applications or limited algorithm comparisons without systematic assessment across multiple performance criteria.

While \cite{tavakkoli2024model} compares a model-free DDPG controller against a classical LQI controller for non-minimum phase systems under various conditions, their analysis is limited to a single DRL algorithm and does not explore the full spectrum of control challenges present in diverse benchmark problems. Similarly, \cite{dutta2022survey} evaluates multiple actor-critic methods but focuses primarily on process nonlinearities without comprehensive analysis of control signal characteristics and robustness metrics that are critical for practical implementation.

Other notable contributions include \cite{dutta2023reinforcement}'s application of DDPG to nonlinear process control, and \cite{joshi2023tasac}'s twin-actor framework for batch processes. However, these studies typically evaluate performance using a narrow set of metrics, often excluding important considerations such as control effort, actuator constraints, and response to uncertainties.

Our work addresses this research gap by providing a systematic, control-oriented evaluation of four leading model-free DRL algorithms across multiple benchmark problems using a comprehensive set of standardised performance criteria. Unlike previous studies that evaluate algorithms using generic reinforcement learning metrics or limited control performance indicators, our approach presents a unified framework for assessment that considers a wide spectrum of characteristics relevant to control engineers.

\begin{table*}[!htbp]
  \centering
  \caption{Overview of Reinforcement Learning Applications in Control}
  \label{tab:rl_process_control}
  \setlength{\tabcolsep}{3pt}
  \small
  \begin{tabularx}{\textwidth}{l >{\centering\arraybackslash}X >{\centering\arraybackslash}X >{\centering\arraybackslash}X >{\centering\arraybackslash}X >{\centering\arraybackslash}X}
    \toprule
    \textbf{Ref.} & \textbf{Application}
      & \textbf{Solution/Algorithm(s)}
      & \textbf{States} 
      & \textbf{Reward Function} & \textbf{Evaluation Metrics} \\
    \midrule
    
    \cite{dutta2023reinforcement}
      & Nonlinear process control (CSTR, pH)
      & DDPG
      & Partial system state
      &
      A $ \beta$ -weighted composite reward, minimizing absolute tracking error and control effort variation
      & Integral Squared Error(ISE), Root Mean Squared Error (RMSE) \\
      
    \cite{joshi2023tasac}
      & Batch process control
      & SAC, DDPG
      & Partial system state
      & Absolute tracking error scaled by time
      & Setpoint tracking error, Disturbance rejection performance \\
      
    \cite{spielberg2019toward}
      & Self-driving industrial processes (HVAC)
      & DPG
      & Partial system state
      & Negative sum of absolute errors (SAE) 
      & Integral Absolute error in case studies \\
      
    \cite{bao2021deep}
      & Learning performance in process control
      & DDACP
      & Partial system state
      & Weighted quadratic performance index balancing tracking errors with control
      & Average Returns  \\
      
    \cite{dutta2022survey}
      & Survey of actor-critic methods
      & Comparison of DDPG, TD3, SAC, PPO, TRPO
      & Partial system state
      & Sparse reward on $\beta$ -weighted performance index
      & Integral Absolute Error (IAE), Settling time  \\
      
    \cite{oh2024quantitative}
      & RL vs.\ MPC for chemical processes
      & DDPG, TD3, SAC
      & Partial system state
      & Negative of the MPC-style stage cost
      & Mean and standard deviation of the cost over multiple Monte-Carlo runs \\
      
    \cite{tavakkoli2024model}
      & Setpoint tracking of non-minimum phase systems
      & DDPG
      & Partial system state
      & Negative LQI cost function
      & IAE; Overshoot; Settling time \\
      
    \cite{tang2025DRLsuccesssurvey}
      & Robotics
      & DDPG, PPO, SAC, TRPO, QR-DQN
      & Partial system state
      & Task-specific rewards
      & Success rate; Sample efficiency \\
    
    \bottomrule
  \end{tabularx}
\end{table*}

\section{The Analysis Framework}

Regardless of the design methodology, for a controller to become operational successfully in the real world, three main aspects of metrics must be satisfied \cite{sarhadi2025standard}: 1) setpoint tracking performance, 2) control signal feasibility or energy consumption, and 3) robustness. During the design stage, controllers are typically developed with specific objectives that may target one or more metrics within these aspects, as reflected in the cost function, stability analysis, or overall control objective. However, once implementation begins, the controller must rigorously satisfy all these aspects; otherwise, re-tuning or redesign will be required, undermining sustainability in the design process. The final consideration is real-time implementability, which no longer poses a significant challenge thanks to the availability of sufficient computational power.

Moreover, the controller should be either designed or at least tested against operational challenges encountered in real-world conditions, such as disturbances and noise, actuation constraints, dynamic limitations including non-minimum phase behaviour or instability, and model uncertainties. Even a model-free controller is typically developed within a simulated environment before transitioning to the real world, in order to reduce the number of required trials and to improve safety and cost-effectiveness, factors that are often jointly considered. These considerations are not only essential for control design but also extend to higher-level functions such as decision-making in autonomous systems. Such challenges are well addressed in classical control, owing to the availability of tools such as frequency domain techniques. However, they are less thoroughly handled in nonlinear systems. As a result, the capabilities and performance boundaries of emerging DRL-based controllers require further investigation.    
Our evaluation methodology is grounded in a performance assessment across the three fundamental categories outlined above, employing the following classes of metrics to meticulously analyse system behaviour:

\begin{enumerate}
\item \textbf{Tracking Metrics}: These quantify how accurately the controller follows the reference signal, including both transient behaviour and steady-state performance, as well as its ability to reject disturbances. Metrics include rise time ($t_r$), maximum overshoot ($M_p$) (or undershoot, $M_u$, where applicable), settling time ($t_s$), steady-state error ($e_{ss}$), integral of squared error (ISE), and integral of time-weighted absolute error (ITAE).

\item \textbf{Control Signal Quality}: These measure characteristics of the control signal that directly affect energy consumption, actuator wear, and implementation feasibility. The criteria in this category include the integral of absolute control effort (IACE), the integral of absolute control effort rate (IACER), and the maximum control value ($u_{\text{max}}$).

\item \textbf{Robustness Characteristics}: These assess the controller's ability to maintain stability and performance in the presence of uncertainties, disturbances, and parameter variations. Relevant metrics include gain margin (GM) and delay margin (DM).
\end{enumerate}

This is based on the understanding that a controller achieving perfect tracking, yet requiring excessive control effort or lacking robustness to uncertainties, holds limited practical value. Table \ref{tab:rl_process_control} presents a comparison of existing research in DRL control analysis, highlighting limitations in evaluation scope. The evaluation criteria are summarised in Table \ref{tab:performance_criteria}, and detailed definitions can be found in \cite{sarhadi2025standard}. Most studies in this field either offer only qualitative assessments \cite{bucsoniu2018RLcomparison}, rely on visual comparisons, or restrict their evaluation to tracking metrics such as the Integral of Absolute Error (IAE).

To address these limitations, our analysis approach encompasses all three performance categories, using the set of metrics specified in Table \ref{tab:performance_criteria}. A unified reward function is employed for all DRL approaches as well as for the LQI controller, facilitating a fair comparison between two distinct optimal control paradigms: a model-free, AI-based solution versus a classical optimal controller with full knowledge of the system model. All controllers are implemented using consistent hyperparameters, network architectures, and training protocols wherever applicable.

Four diverse systems are selected: non-minimum phase (NMP), two-mass spring (TMS), AUV (nonlinear), and Crazyflie quadrotor. For each benchmark problem, we conduct two primary sets of tests: nominal performance tests examining step response under ideal conditions to assess basic tracking performance, and robustness evaluation tests including disturbances and noise to assess disturbance rejection capabilities.

The results are quantified using a comprehensive performance metrics detailed in Table~\ref{tab:performance_criteria}, providing a systematic evaluation framework that encompasses both transient response characteristics and practical implementation considerations. Each metric addresses specific aspects of control performance relevant to industrial applications, enabling objective comparison across fundamentally different algorithmic approaches. By implementing this rigorous evaluation methodology, our study establishes a bridge between theoretical reinforcement learning research and practical control engineering applications, providing quantitative guidance for algorithm selection based on specific system requirements and operational constraints.

\begin{table}[b!]
 \caption{Performance Criteria for Comprehensive Controller Evaluation}
 \label{tab:performance_criteria}
 \centering
 \footnotesize
 \setlength{\tabcolsep}{4pt}
 \begin{tabular}{@{} lccc @{}}
   \toprule
   \textbf{Metric}
     & \textbf{Mathematical Definition}
     & \textbf{Engineering Significance}
     & \textbf{Unit} \\
   \midrule
   $t_r$      & Rise time (10\% to 90\% of reference)   & Speed of initial response                & s   \\
   $M_p$      & Maximum percentage overshoot            & Peak transient above reference           & \%  \\
   $t_s$      & Settling time to 2\% of final value     & Time to reach steady-state               & s   \\
   $e_{\mathrm{ss}}$
              & $\displaystyle\lim_{t\to\infty}\lvert r(t)-y(t)\rvert$
                                                        & Steady-state tracking accuracy           & --   \\
   ISE        & $\displaystyle\int_0^T e^2(t)\,\mathrm{d}t$        
                                                        & Penalises large tracking errors          & --  \\
   ITAE       & $\displaystyle\int_0^T t\,\lvert e(t)\rvert\,\mathrm{d}t$
                                                        & Emphasises persistent late errors        & --  \\
   IACE       & $\displaystyle\int_0^T \lvert u(t)\rvert\,\mathrm{d}t$
                                                        & Total control effort and energy          & --  \\
   IACER      & $\displaystyle\int_0^T \bigl\lvert\tfrac{\mathrm{d}u}{\mathrm{d}t}\bigr\rvert\,\mathrm{d}t$
                                                        & Control signal smoothness                & --  \\
   $u_{\max}$ & $\displaystyle\max_t\lvert u(t)\rvert$  & Peak control demand                      & --  \\
   GM         & $20\log_{10}\bigl(\tfrac{1}{|G(j\omega_{pc})|}\bigr)$ & Stability robustness to gain variations & dB  \\
   DM         & $\displaystyle\frac{\text{PM}}{\omega_{cg}}$     & Robustness to time delays               & s   \\
   \bottomrule
 \end{tabular}
\end{table}

The selected metrics evaluate controller performance across multiple dimensions. Time-domain metrics ($t_r$, $M_p$, $t_s$, $e_{\mathrm{ss}}$) characterise transient and steady-state response quality, where $r(t)$ represents the reference signal and $y(t)$ denotes the system output. Integral performance indices (ISE, ITAE) provide weighted assessments of tracking accuracy over the evaluation period $T$, with $e(t) = r(t) - y(t)$ representing the tracking error. Control effort metrics (IACE, IACER, $u_{\max}$) quantify practical implementation costs where $u(t)$ represents the control signal, encompassing energy consumption, actuator wear, and saturation risks.

Stability margins provide critical robustness assessments: gain margin (GM) measures the allowable gain increase before instability, calculated at the phase crossover frequency $\omega_{pc}$ where the open-loop system $G(j\omega)$ has $-180 ^\circ $ phase. Delay margin (DM) quantifies robustness to time delays, derived from the phase margin (PM) measured in degrees and the gain crossover frequency $\omega_{cg}$ where $|G(j\omega_{cg})| = 1$. 

This multi-faceted evaluation approach enables identification of fundamental performance trade-offs inherent in different control strategies, providing engineers with quantitative guidance for selecting algorithms based on application-specific priorities such as speed, accuracy, energy efficiency, or robustness requirements.

\section{Simulation Studies}
This section presents simulation results and analysis for our benchmark problems, comparing the performance of DDPG, TD3, PPO, and TD-MPC2 algorithms across multiple scenarios and evaluation criteria.
\subsection{Simulation Setup}
The simulation environment was implemented in Python using PyTorch for the deep learning components and Control Systems Library for the dynamic systems simulation. For reproducibility and consistency across benchmarks, we maintained standardised neural network architectures and optimisation procedures if possible whilst adapting hyperparameters for each specific control task. Table~\ref{tab:drl_parameters} presents the hyperparameter configuration for the Non-minimum phase system (NMPs) benchmark, which follows the same structure used across all experiments. Comprehensive hyperparameter settings for all other benchmarks are available in our public repository\footnote{\href{https://github.com/Klins101/Performance-Analysis-of-Deep-Reinforcement-Learning-for-Applied-Control-Applications.git}{\textcolor{blue}{https://github.com/Klins101/Performance-Analysis-of-Deep-Reinforcement-Learning-for-Applied-Control-Applications.git}}}. 

\subsection{Non-Minimum Phase CSTR plane}

The first system is a Continuous Stirred-Tank Reactor (CSTR) process control benchmark with the following transfer function \cite{bequette2003processcstr}:
\begin{equation}
\large
G(s) = \frac{-1.135s + 3.199}{s^2 + 4.719s + 5.47}
\end{equation}

The presence of the unstable zero introduces fundamental performance limitations and undershoot behaviour that significantly challenges control system design \cite{aastrom2000limitations, sarhadi2025standard}. Despite these challenges being documented in classical control theory, the efficacy of deep reinforcement learning approaches in managing such dynamics remains relatively unexplored \cite{tavakkoli2024model}. Algorithms are trained with parameters in Table~\ref{tab:drl_parameters}.

\begin{table}[t!]
\caption{DRL Implementation Parameters}
\label{tab:drl_parameters}
\centering
\small 
\setlength{\tabcolsep}{4pt} 
\renewcommand{\arraystretch}{1.1} 
\begin{tabular}{p{2.8cm}cccc} 
\hline
\textbf{Parameter} & \textbf{DDPG} & \textbf{TD3} & \textbf{PPO} & \textbf{TD-MPC2} \\
\hline
Total timesteps & 200,000 & 200,000 & 200,000 & 200,000 \\
Batch size & 256 & 256 & 64 & 256 \\
Discount factor ($\gamma$) & 0.99 & 0.99 & 0.99 & 0.99 \\
Target update ($\tau$) & 0.005 & 0.005 & N/A & 0.001 \\ 
Actor lr & 1e-4 & 1e-4 & 1e-4 & 1e-4  \\ 
Critic lr & 1e-3 & 1e-3 & 1e-3 & 1e-3  \\ 
Hidden dim & 64 & 64 & 64 & 128 \\
Actor net & [64, 64] & [64, 64] & [64, 64] & [64, 64] \\
Critic net & [64, 64] & [64, 64] x2 & [64, 64] & [64, 64]  \\ 
Expl noise & OU Noise & Gaussian & N/A & Gaussian \\ 
Policy noise & N/A & 0.2 & N/A & N/A \\
Noise clip & N/A & 0.5 & N/A & N/A \\
Policy delay & 1 & 2 & N/A & N/A \\
Buffer size & 100,000 & 100,000 & N/A & 100,000 \\
Clip param & N/A & N/A & 0.2 & N/A \\
Sim dt & 0.1 s & 0.1 s & 0.1 s & 0.1 s \\ 
Ep length & 25 s & 25 s & 25 s & 25 s \\ 
Activation & ReLU & ReLU & ReLU & ReLU \\
Output act & Tanh & Tanh & Tanh & Tanh \\ 
\hline

\end{tabular}
\end{table}

All controllers were implemented using the same environment, which provides a consistent interface for training and evaluating the different algorithms across all benchmark problems. The environment encapsulates system dynamics, reference generation, disturbance application, and performance measurement. For each benchmark system, the appropriate state-space model was integrated within this environment framework.

\subsubsection{Performance Results}
\begin{figure}[t!]
\centering
\includegraphics[width=1.0\linewidth]{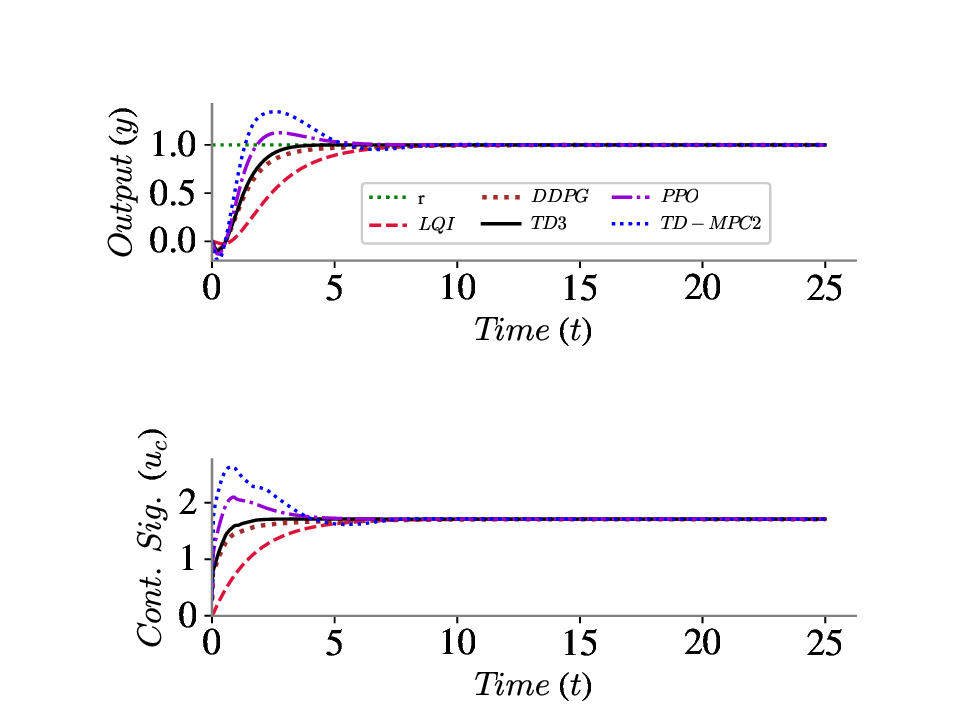}
\caption{Results for the NMP system in ideal conditions}
\label{fig:nonminphase_response}
\end{figure}

Figure~\ref{fig:nonminphase_response} shows the simulation results for the non-minimum phase system. Table~\ref{tab:nonminphase_results} presents the performance metrics for the non-minimum phase CSTR system. All controllers exhibit the characteristic inverse response, with initial movement opposite to the reference direction before convergence.

\begin{table}[t!]
\caption{Performance Results for Non-Minimum Phase System}
\label{tab:nonminphase_results}
\centering
\begin{tabular}{lccccc}
\hline
\textbf{Metric} & \textbf{LQI} & \textbf{DDPG} & \textbf{TD3} & \textbf{PPO} & \textbf{TD-MPC2}\\
\hline
$t_r$ (s) & 3.90 & 2.30 & 1.70 & 0.90 & 0.60 \\
$M_p$ (\%) & 0.00 & 0.00 & 0.10 & 12.60 & 34.50 \\
$t_s$ (s) & 7.70 & 6.10 & 3.30 & 5.50 & 8.10 \\
$e_{\text{ss}}$ & 0.00 & 0.00 & 0.00 & 0.00 & 0.00 \\
ISE & 2.00 & 1.20 & 1.10 & 1.00 & 1.20 \\
ITAE & 5.60 & 2.80 & 1.30 & 1.80 & 3.80 \\
IACE & 39.90 & 41.70 & 42.30 & 43.40 & 44.40 \\
IACER & 1.65 & 1.35 & 1.40 & 2.20 & 3.50 \\
$u_{\max}$ & 1.65 & 1.73 & 1.81 & 2.10 & 2.64 \\
GM & 40.00 & 2.14 & 2.51 & 1.90  & 1.83       \\
DM & 1.70 & 1.61 & 1.58 & 0.90  & 0.7       \\
\hline
\end{tabular}
\end{table}

TD-MPC2 achieves the fastest rise time (0.60s) but with substantial overshoot (34.50\%) and the highest control effort ($u_{\max} = 2.64$). PPO demonstrates rapid response (0.90s) with moderate overshoot (12.60\%) and the best ISE performance (1.00) among all controllers. TD3 delivers balanced performance with minimal overshoot (0.10\%), excellent settling time (3.30s), and superior ITAE (1.30), indicating effective error minimisation over time. DDPG shows conservative behaviour similar to TD3 but with slower dynamics.

The baseline LQI controller exhibits zero overshoot but significantly slower response ($t_r = 3.90$s, $t_s = 7.70$s) and requires the lowest control effort (IACE = 39.90). All DRL algorithms demand higher control activity than LQI, with IACE values ranging from 41.70 to 44.40, reflecting the trade-off between response speed and actuator demands.

Robustness analysis reveals LQI's superior gain margin (40.00) compared to DRL algorithms (1.83-2.51). Among DRL controllers, TD3 achieves the highest gain margin (2.51) whilst DDPG provides the best delay margin (1.61), suggesting varying robustness characteristics across algorithms.

\subsubsection{Robustness Under Perturbed Conditions}
\begin{figure}[t!]
\centering
\includegraphics[width=1.0\linewidth]{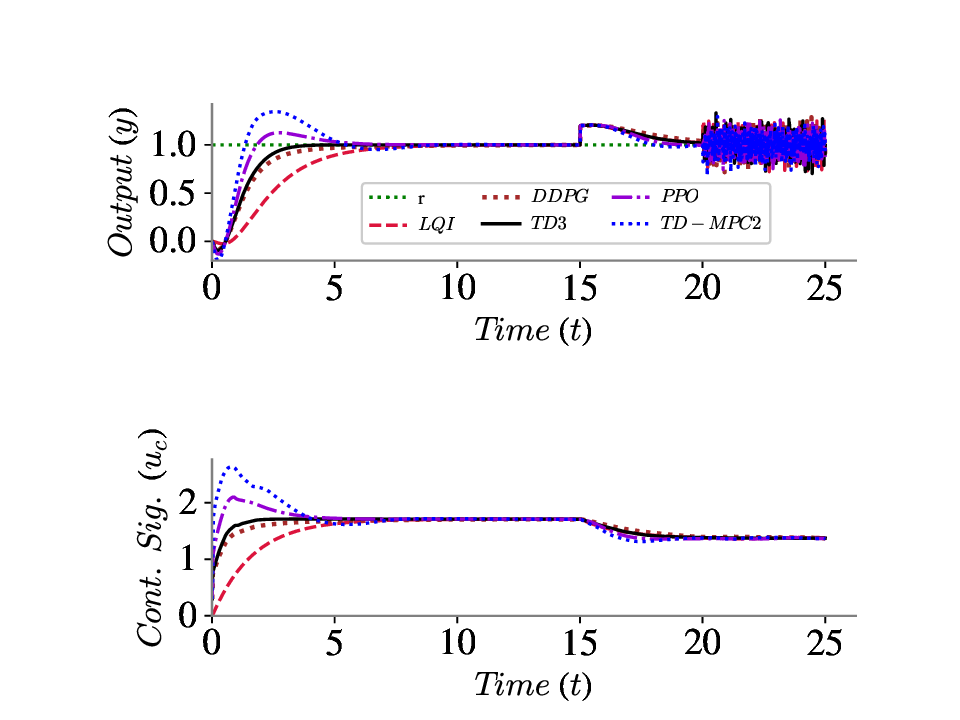}
\caption{Results for NMP in the perturbed test}
\label{fig:nonminphase_disturbance}
\end{figure}

Under perturbed conditions with step disturbance (0.20 at $t = 15$s) and measurement noise ($\sigma = 0.20$ at $t = 20$s) as shown in Figure~\ref{fig:nonminphase_disturbance}, all controllers maintain stability whilst exhibiting distinct recovery characteristics. DRL algorithms demonstrate faster disturbance recovery than LQI, with TD3 returning to within 2\% of reference in approximately 3.50s compared to LQI's 6.00s. When noise is introduced, TD3 maintains the tightest output bounds whilst TD-MPC2 shows increased sensitivity with larger variations.

The results highlight fundamental trade-offs in non-minimum phase control: aggressive controllers (TD/-MPC2, PPO) achieve faster response at the cost of overshoot and control effort, whilst conservative approaches (TD3, DDPG, LQI) prioritise stability. TD3 emerges as the most balanced solution, combining reasonable speed, minimal overshoot, and robust disturbance rejection, critical attributes for practical implementation in chemical process control where both performance and stability are essential.

\subsubsection{Analysis of Performance Trade-offs}
The results highlight fundamental trade-offs in controlling non-minimum phase systems. A clear inverse relationship between rise time and overshoot is evident: TD-MPC2 and PPO achieve the fastest responses (0.60s, 0.90s) but with significant overshoots (34.50\%, 12.60\%), whilst TD3 and DDPG maintain minimal overshoot at the cost of slower rise times.

PPO achieves the best tracking performance (ISE = 1.00) among all controllers, though TD3 demonstrates superior time-weighted error minimisation (ITAE = 1.30) and the fastest settling time (3.30s) with negligible overshoot (0.10\%). This combination suggests TD3's control strategy effectively balances transient response with long-term accuracy.

Control effort analysis reveals a consistent pattern: LQI requires the lowest effort (IACE = 39.90) but delivers the slowest response, whilst DRL algorithms trade increased control activity (IACE: 41.70-44.40) for improved dynamics. Among DRL controllers, TD3 achieves the highest gain margin (2.51) whilst DDPG provides the best delay margin (1.61), indicating different robustness properties.

The comparative analysis indicates that controller selection depends on application priorities: PPO for minimal tracking error, TD3 for balanced performance with robustness, TD-MPC2 for fastest response, or LQI for minimal control effort. For most practical applications, TD3's combination of reasonable speed, minimal overshoot, and robust margins provides the most versatile solution.

\subsection{Two Mass Spring Benchmark Problem}
The second benchmark considered is the infamous ACC 1992 benchmark problem \cite{wie1992benchmarkacc}, Two Mass Spring System(TMS). First introduced in 1992 as a touchstone benchmark in testing control systems, it is important because of the non-collocated actuator and sensor dynamics, making it a prevalent benchmark in control literature \cite{wie1992benchmark}. Its dynamics are represented in state-space form as:

\begin{equation}
\large
\begin{bmatrix}
\dot{x}_1 \\
\dot{x}_2 \\
\dot{x}_3 \\
\dot{x}_4
\end{bmatrix}
=
\begin{bmatrix}
0 & 0 & 1 & 0 \\
0 & 0 & 0 & 1 \\
-\tfrac{k}{m_1} & \tfrac{k}{m_1} & 0 & 0 \\
\tfrac{k}{m_2} & -\tfrac{k}{m_2} & 0 & 0 \\
\end{bmatrix}
\begin{bmatrix}
x_1 \\
x_2 \\
x_3 \\
x_4
\end{bmatrix}
+
\begin{bmatrix}
0 \\
0 \\
\tfrac{1}{m_1} \\
0
\end{bmatrix}u
\end{equation}

where $x_1$ and $x_2$ are the positions of body 1 and body 2, respectively; $x_3$ and $x_4$ are the velocities; $u$ is the control input acting on body 1.

For our nominal system, we set $m_1 = m_2 = 1$ and $k = 1$ with appropriate units. It's challenging and used as a main benchmark in numerous control studies

\subsubsection{Performance Results}
\begin{figure}[t!]
\centering
\includegraphics[width=1.0\linewidth]{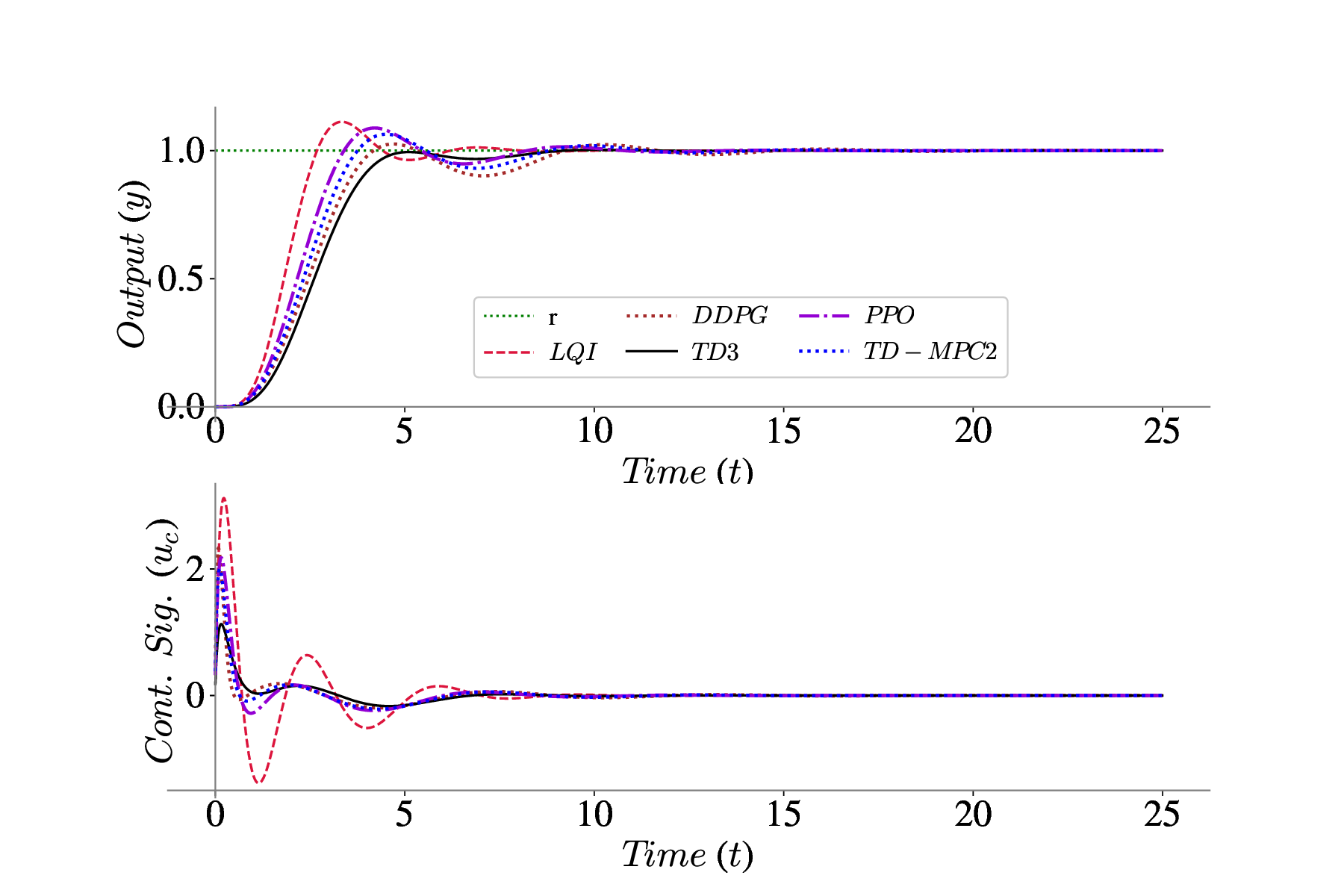}
\caption{Step response comparison for TMS system showing nominal performance characteristics}
\label{fig:tms_response}
\end{figure}

The nominal step response results presented in Fig.~\ref{fig:tms_response} reveal distinctly different performance characteristics among the evaluated controllers. The quantitative analysis confirms significant algorithmic differences in handling the system's oscillatory dynamics and resonant modes.

Table~\ref{tab:tms_performance_nominal} presents the nominal performance metrics. LQI achieves the fastest rise time (1.40s) with moderate overshoot (11.20\%), leveraging model-based design for precise dynamics exploitation. However, it requires the highest control effort (IACE = 15.18) whilst maintaining superior control smoothness (IACER = 45.86).

Among DRL algorithms, TD3 demonstrates precision with minimal overshoot (0.30\%) and the lowest maximum control effort ($u_{\max} = 1.18$). Its twin-critic architecture effectively mitigates overestimation bias, achieving the smoothest control signals (IACER = 10.91) at the cost of slower rise time (2.50s). PPO provides balanced performance with reasonable transient response (1.90s rise time, 8.80\% overshoot) and moderate control effort, whilst DDPG and TD-MPC2 exhibit more conservative behaviours.

Tracking performance analysis reveals LQI's superiority with the lowest ISE (6.04) and ITAE (10.07), reflecting optimal design benefits. DRL algorithms show higher tracking errors (ISE: 7.13-8.56) but significantly lower control efforts (IACE: 4.52-6.79), highlighting the fundamental trade-off between tracking precision and actuator demands.

\subsubsection{Robustness Under Perturbed Conditions}

\begin{figure}[t!]
\centering
\includegraphics[width=1.0\linewidth]{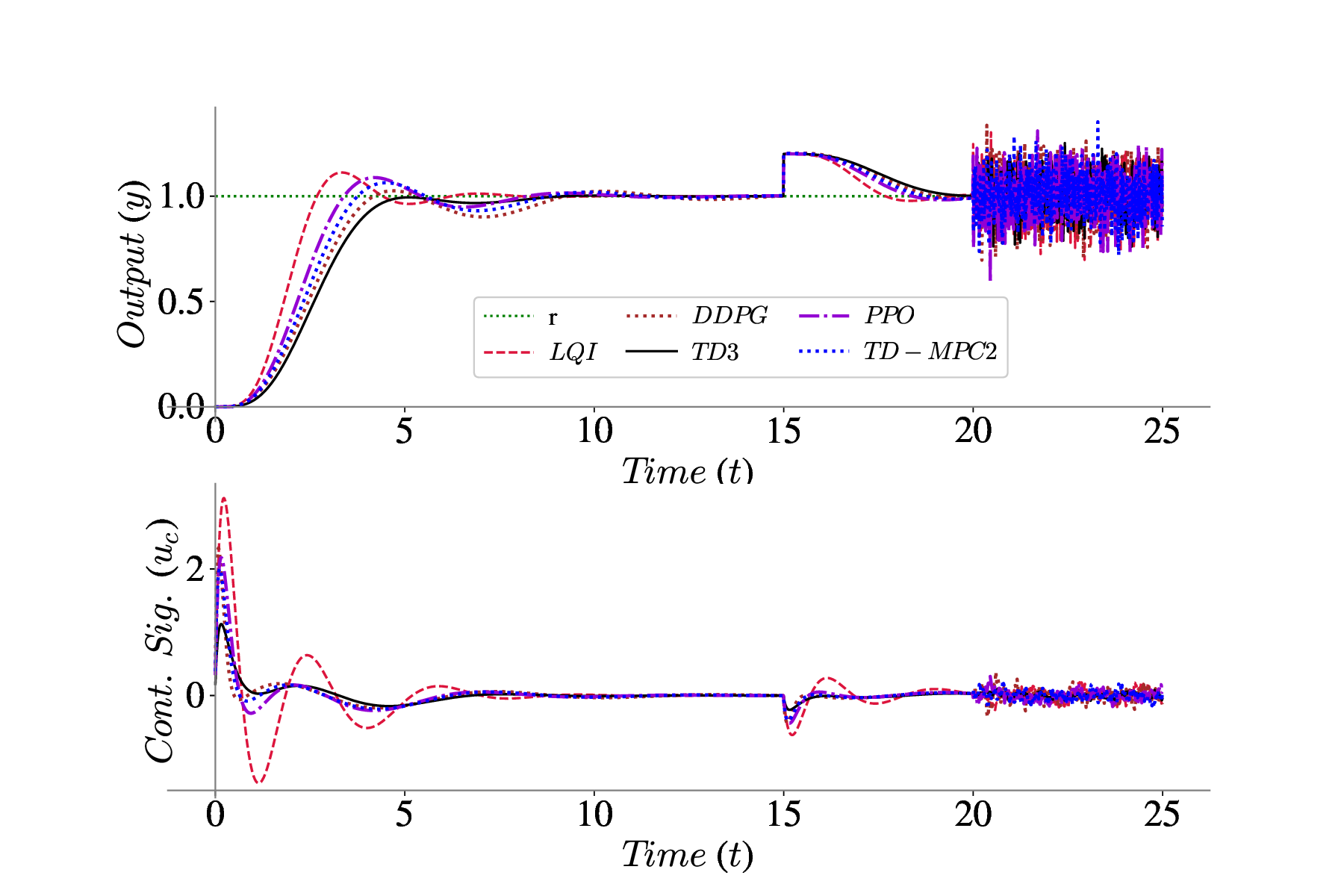}
\caption{TMS system performance under perturbed conditions with disturbances and measurement noise}
\label{fig:mass_spring}
\end{figure}

Under perturbed conditions incorporating white noise and external disturbances (Fig.~\ref{fig:mass_spring}), the relative performance characteristics reveal critical insights into algorithm robustness and practical applicability. The dramatic shift in performance rankings under uncertainty demonstrates the importance of evaluating controllers beyond nominal conditions.

Table~\ref{tab:tms_performance_perturbed} quantifies performance under disturbances and measurement noise. All controllers experience degradation, with ISE values increasing by 7.6-10.5\% for LQI and DRL algorithms. LQI maintains the best tracking (ISE = 6.5) but requires increased control effort (IACE rising from 15.18 to 19.6).

TD3 exhibits superior control smoothness under perturbations (IACER = 43.3), significantly outperforming other algorithms. This robustness, combined with the lowest control effort (IACE = 5.9), validates the twin-critic architecture's effectiveness for uncertain environments. PPO demonstrates balanced robustness with moderate degradation across all metrics, whilst DDPG shows the highest sensitivity to disturbances.

\subsubsection{Analysis of Performance Trade-offs}

\begin{table}[h!]
\centering
\caption{Performance metrics for TMS system under nominal conditions}
\label{tab:tms_performance_nominal}
\setlength{\tabcolsep}{6pt}
\renewcommand{\arraystretch}{1.0}
\begin{tabular}{lccccc}
\hline
\textbf{Metric}    & \textbf{LQI} & \textbf{DDPG} & \textbf{TD3} & \textbf{PPO} & \textbf{TD-MPC2}\\
\hline
$t_r$ (s) & 1.40 & 2.30 & 2.50 & 1.90 & 2.10 \\
$M_p$ (\%) & 11.20 & 2.60 & 0.30 & 8.80 & 6.40 \\
$t_s$ (s) & 5.70 & 10.70 & 7.90 & 7.70 & 8.30 \\
$e_{\mathrm{ss}}$ & 0.00 & 0.00 & 0.00 & 0.00 & 0.00 \\
ISE & 6.04 & 7.92 & 8.56 & 7.13 & 7.58 \\
ITAE & 10.07 & 24.91 & 18.24 & 16.02 & 19.76 \\
IACE & 15.18 & 5.73 & 4.52 & 6.79 & 5.84 \\
IACER & 45.86 & 21.34 & 10.91 & 22.37 & 19.15 \\
$u_{\max}$ & 3.14 & 2.32 & 1.18 & 2.26 & 2.03 \\
GM (dB) & 21.15 & 1.43 & 2.85 & 1.95 & 2.45 \\
DM (s) & 0.17 & 0.25 & 0.52 & 0.35 & 0.48 \\
\hline
\end{tabular}
\end{table}

\begin{table}[h!]
\centering
\caption{Performance metrics for TMS system under perturbed conditions}
\label{tab:tms_performance_perturbed}
\setlength{\tabcolsep}{6pt}
\renewcommand{\arraystretch}{1.0}
\begin{tabular}{lccccc}
\hline
\textbf{Metric}    & \textbf{LQI} & \textbf{DDPG} & \textbf{TD3} & \textbf{PPO} & \textbf{TD-MPC2}\\
\hline
ISE & 6.5 & 8.5 & 9.0 & 7.5 & 8.0 \\
ITAE & 70.5 & 93.7 & 89.3 & 81.0 & 86.0 \\
IACE & 19.6 & 8.1 & 5.9 & 8.9 & 7.8 \\
IACER & 106.2 & 109.1 & 43.3 & 80.7 & 86.5 \\
\hline
\end{tabular}
\end{table}

The TMS benchmark reveals distinct control philosophies: LQI prioritises tracking accuracy through high control effort, achieving superior nominal performance but requiring substantial actuator activity. TD3 optimises control economy and smoothness, sacrificing response speed for precision and robustness critical attributes for mechanical systems with actuator constraints.

Robustness analysis confirms LQI's superior gain margin (21.15 dB) compared to DRL algorithms (1.43-2.85 dB), whilst TD3 achieves the best delay margin (0.52s) among DRL controllers. These metrics suggest complementary robustness properties: classical control excels in parameter variations whilst TD3 handles time delays more effectively.

The results demonstrate that controller selection depends on application priorities. LQI suits applications demanding precise tracking with available actuator capacity, TD3 excels where control smoothness and economy are paramount, whilst PPO offers reliable balanced performance. The significant performance variations under uncertainty underscore the importance of robustness evaluation beyond nominal conditions for practical deployments.

\subsection{Nonlinear Autonomous Underwater Vehicle (AUV)}
The AUV benchmark represents the complex nonlinear yaw dynamics of the REMUS AUV \cite{prestero2001development} incorporating realistic hydrodynamic effects including quadratic damping, velocity-dependent coefficients, and cross-coupling between sway and yaw motions. The system dynamics are governed by:

\begin{figure*}[h!]

\begin{equation}
\displaystyle
\begin{cases}
(m - Y_{\dot{v}}) \dot{v} + (m x_G - Y_{\dot{r}}) \dot{r} = \left( Y_{v|v|} |v| + Y_{uv} u_o \right) v + \left( Y_{r|r|} |r| + (Y_{ur} - m) u_o \right) r + Y_{uu \delta_r} u_o^2 \delta_r \\
(m x_G - N_{\dot{v}}) \dot{v} + (I_z - N_{\dot{r}}) \dot{r} = \left( N_{v|v|} |v| + N_{uv} u_o \right) v + \left( N_{r|r|} |r| + (N_{ur} - m x_G) u_o \right) r + N_{uu \delta_r} u_o^2 \delta_r \\
\dot{\psi} = r
\end{cases}
\label{eq:AUV_dynamics}
\end{equation}

\end{figure*}

All parameters and variables of this model are defined in Table \ref{table:auv_variables} and are based on empirically validated REMUS AUV from \cite{prestero2001development}. This benchmark represents a challenging control problem on the realistic nonlinear model of an AUV with inherent non-BIBO stability, added mass effects, and significant nonlinearities that require active stabilisation, making it representative of real-world maritime control scenarios. To comprehensively evaluate controller performance, we incorporate realistic operational constraints including actuator amplitude and rate saturations ($|\delta_r| \leq 20$, $|\dot{\delta}_r| \leq 30$), unmodelled dynamics representing worst-case uncertainties, and environmental disturbances typical of maritime operations. These conditions test the algorithms' ability to handle the practical limitations encountered in autonomous underwater vehicle deployment beyond idealised simulation scenarios.

\begin{table*}
\caption{Model parameters and state variables used in AUV dynamics}
\renewcommand{\arraystretch}{1.1}
\label{table:auv_variables}
\centering
\footnotesize
\begin{tabular}{m{15mm} m{30mm} m{10mm} m{15mm} m{30mm} m{10mm}}

\toprule
\textbf{Param./Var.} & \textbf{Description} & \textbf{Value} & \textbf{Param./Var.} & \textbf{Description} & \textbf{Value} \\
\midrule

$m$ (kg) & Vehicle mass & 30.50 & $x_G$ (m) & CG position (long.) & 0.00 \\
$I_z$ (kg.m\textsuperscript{2}) & Yaw inertia & 3.45 & $Y_{\dot v}$ (kg.m/rad) & Added Mass (sway) & -35.50 \\
$Y_{uv}$ (kg/m) & Body lift force and fin lift & -28.60 & $Y_{v|v|}$ (kg/m) & Cross-flow Drag (sway) & -1310.00 \\
$Y_{ur}$ (kg/rad) & Added mass cross term and fin lift & 5.22 & $Y_{r|r|}$ (kg.m\textsuperscript{2}/rad) & Cross-flow drag (yaw) & 0.63 \\
$Y_{uu\delta_r}$ (kg/m.rad) & Fin lift force & 21.37 & $N_{\dot v}$ (kg.m) & Added mass (sway-yaw) & 1.93 \\
$N_{\dot r}$ (kg.m\textsuperscript{2}/rad) & Added mass (yaw) & -4.88 & $N_{v|v|}$ (kg) & Cross-flow drag (sway) & -3.18 \\
$N_{r|r|}$ (kg.m\textsuperscript{2}/rad\textsuperscript{2}) & Cross-flow drag (yaw) & -94.00 & $N_{uu\delta_r}$ (kg/rad) & Body and fin lift and munk moment & -17.50 \\
$N_{uv}$ (kg) & Fin lift moment (surge-sway) & -24.00 & $N_{ur}$ (kg.m/rad) & Added mass cross term and fin lift & -2.00 \\
$u_0$ (m/s) & Forward speed (constant) & 1.50 & $\delta_r$ (rad) & Rudder angle (input) & NA \\
$v$ (m/s) & Sway velocity (state) & NA & $r$ (rad/s) & Yaw rate (state) & NA \\
$\psi$ (rad) & Heading angle (state) & NA & & & \\
\bottomrule
\end{tabular}
\end{table*}

\subsubsection{Performance Analysis}
\begin{figure}[t!]
\centering
\includegraphics[width=1.0\linewidth]{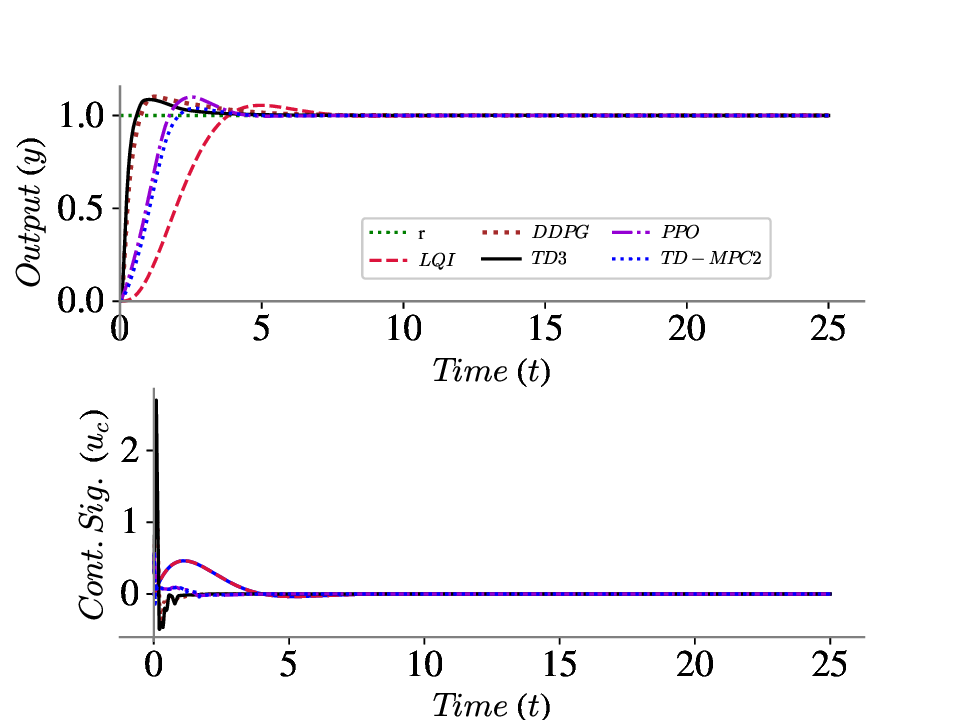}
\caption{Step response comparison for nonlinear AUV system}
\label{fig:auv_response}
\end{figure}

Figure~\ref{fig:auv_response} reveals distinctly different performance characteristics across all evaluated algorithms, with significant variations in transient response and control aggressiveness. The quantitative results demonstrate the algorithms' varying degrees of adaptation to the complex hydrodynamic dynamics inherent in marine vehicle control.

TD3 exhibits the most aggressive response characteristics with the fastest rise time (0.3s) and rapid convergence, though this performance comes at the cost of significant overshoot (8.6\%) and the highest control effort demands. The algorithm demonstrates exceptional settling performance (2.8s) and superior error minimisation (ISE = 1.7), indicating effective exploitation of the system's control authority despite the challenging nonlinear dynamics.

DDPG achieves similarly rapid response with competitive rise time (0.5s) and substantial overshoot (10.1\%), demonstrating effective handling of the nonlinear dynamics through aggressive control strategies. The algorithm shows excellent tracking performance with low ISE values whilst maintaining reasonable settling characteristics. PPO delivers more conservative performance with moderate rise time (1.3s) and overshoot (9.9\%), reflecting its inherent policy stability mechanisms that provide robustness at the expense of transient aggressiveness.

TD-MPC2 exhibits notably conservative response characteristics with slower rise time (1.4s) compared to other DRL methods, though achieving excellent overshoot control (3.9\%) that approaches the classical baseline. This behaviour reflects the model-based approach's tendency towards cautious control strategies when operating with learned dynamics models, prioritising stability over aggressive performance. LQI provides the most conservative response with minimal overshoot (5.5\%) but significantly slower rise time (2.3s), clearly demonstrating the fundamental limitations of linearised control applied to this inherently nonlinear marine system.

\begin{table}[b!]
\caption{Performance Results for Nonlinear AUV System}
\label{tab:auv_results}
\centering
\setlength{\tabcolsep}{8pt}
\renewcommand{\arraystretch}{1.1}
\begin{tabular}{lccccc}
\hline
\textbf{Metric}    & \textbf{LQI} & \textbf{DDPG} & \textbf{TD3} & \textbf{PPO} & \textbf{TD-MPC2}\\
\hline
$t_r$ (s) & 2.3 & 0.5 & 0.3 & 1.3 & 1.4 \\
$M_p$ (\%) & 5.5 & 10.1 & 8.6 & 9.9 & 3.9 \\
$t_s$ (s) & 6.6 & 4.5 & 2.8 & 3.9 & 3.5 \\
$e_{\text{ss}}$ & 0.00 & 0.00 & 0.00 & 0.00 & 0.00 \\
ISE & 15.2 & 2.1 & 1.7 & 6.5 & 7.4 \\
ITAE & 31.2 & 7.1 & 3.3 & 9.6 & 8.2 \\
IACE & 11.0 & 2.9 & 3.9 & 1.5 & 1.3 \\
IACER & 10.0 & 43.9 & 65.5 & 11.1 & 20.3 \\
$u_{\max}$ & 0.5 & 1.9 & 2.7 & 0.6 & 0.6 \\
GM & 25.30 & 16.20 & 15.85 & 16.45 & 13.32 \\
DM & 0.69 & 0.35 & 0.32 & 0.38 & 0.42 \\
\hline
\end{tabular}
\end{table}

The control effort analysis reveals significant disparities in actuator demands across algorithms. LQI maintains exceptionally smooth control signals with minimal peak demand (0.5), reflecting its optimal design principles but at the cost of sluggish response characteristics. Conversely, DRL algorithms exhibit substantially higher control activity, with TD3 demanding maximum control authority (2.7) and the highest control rate variations (IACER = 65.5), indicating aggressive policy behaviour that may challenge actuator limitations in practical implementations.

\subsubsection{Disturbance and Noise Rejection}
Table~\ref{tab:auv_performance_perturbed} presents controller performance under perturbed conditions incorporating measurement noise and external disturbances. All controllers exhibit minimal performance degradation, with ISE values increasing by factors of 1.04-1.12 compared to nominal conditions. 

LQI shows minimal degradation with ISE increasing from 15.2 to 15.8, whilst maintaining moderate control effort (IACE = 13.2) amongst all controllers. The DRL algorithms demonstrate varying robustness characteristics: TD3 achieves the best tracking performance (ISE = 1.9) despite disturbances, whereas TD-MPC2 exhibits slightly higher degradation with ISE increasing from 7.4 to 7.6. 

\begin{figure}[t!]
\includegraphics[width=1.0\linewidth]{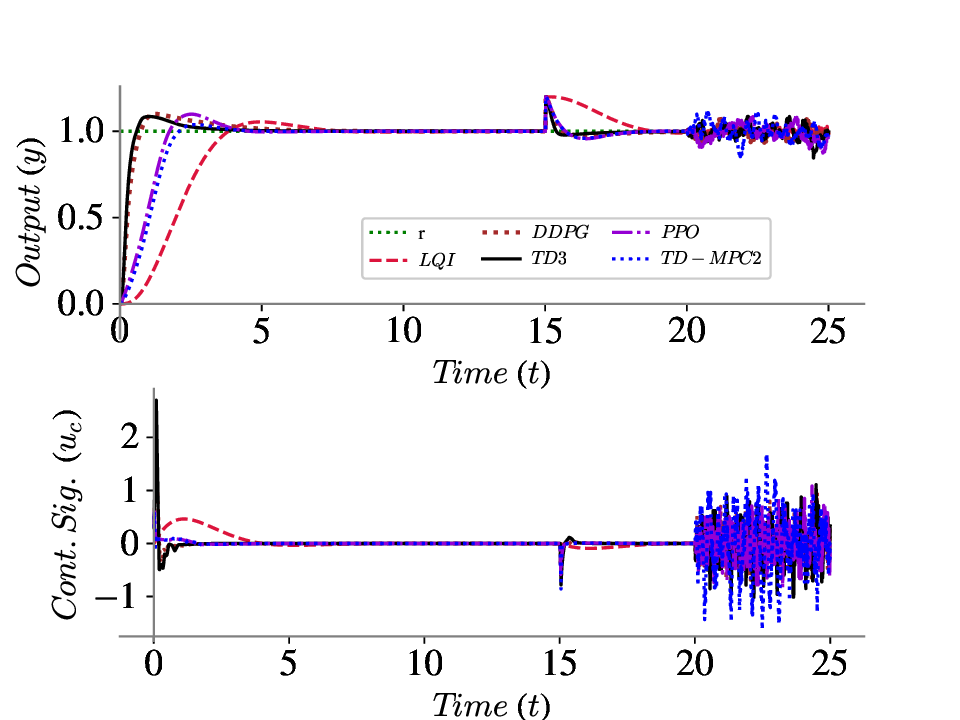}
\caption{AUV performance under perturbed conditions with measurement noise and external disturbances}
\label{fig:auv_noise}
\end{figure}

The control activity metrics reveal significant increases across all algorithms, with IACER values exceeding 140 for most DRL controllers. TD-MPC2 shows the highest control rate activity (IACER = 149.9), indicating aggressive compensation strategies that may challenge actuator bandwidth limitations. PPO maintains moderate performance degradation (ISE = 6.7) with intermediate control effort, suggesting a balanced robustness-performance trade-off. These results as shown in Figure~\ref{fig:auv_noise} highlight the inherent challenge of maintaining smooth control under uncertainty, regardless of the control paradigm employed.

\begin{table}[!t]
\centering
\caption{Performance metrics for AUV system under perturbed conditions}
\label{tab:auv_performance_perturbed}
\setlength{\tabcolsep}{6pt}
\renewcommand{\arraystretch}{1.0}
\begin{tabular}{lccccc}
\hline
\textbf{Metric}    & \textbf{LQI} & \textbf{DDPG} & \textbf{TD3} & \textbf{PPO} & \textbf{TD-MPC2}\\
\hline
ISE & 15.8 & 2.3 & 1.9 & 6.7 & 7.6 \\
ITAE & 107.3 & 52.8 & 48.1 & 72.8 & 71.0 \\
IACE & 13.2 & 15.3 & 20.0 & 14.6 & 25.3 \\
IACER & 12.4 & 141.2 & 143.8 & 142.9 & 149.9 \\
\hline
\end{tabular}
\end{table}

\subsubsection{Performance Evaluation Under Operational Challenges}

To assess controller performance under realistic operational constraints, a comprehensive assessment incorporating actuator saturations, model uncertainties, and external perturbations was conducted. The reference amplitude is increased to 10 to activate actuator saturation limits of $|\delta_r| \leq 20$ and rate constraints of $|\dot{\delta}_r| \leq 30$/s. Model uncertainty is introduced through unmodelled dynamics $G_u(s) = 225/(s^2 + 12s + 225)$, representing Rohrs' example \cite{1104070}, a worst-case scenario known to challenge adaptive controllers. External perturbations include a step disturbance of 2 at $t = 15$s and measurement noise with $\sigma = 0.05$ at $t = 20$s.

\begin{figure}[t!]
\includegraphics[width=1.0\linewidth]{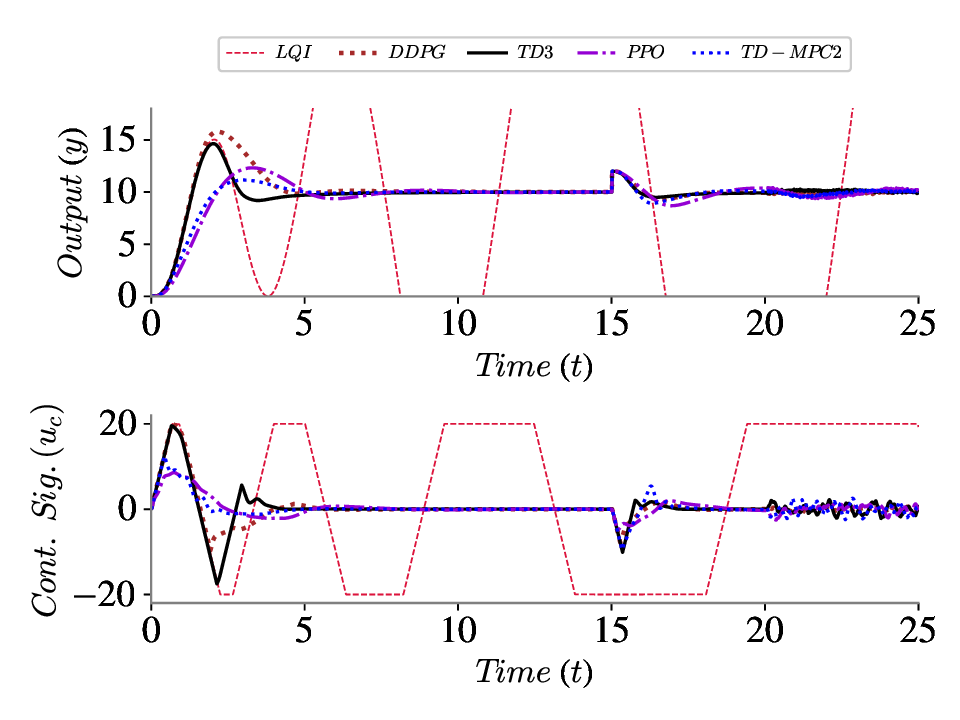}
\caption{Controller performance comparison with operational challenges}
\label{fig:oc}
\end{figure}

Figure~\ref{fig:oc} illustrates the controller responses under these challenging conditions. The LQI controller exhibits sensitivity to the windup phenomenon, resulting in unbounded oscillations. In contrast, DDPG and TD3 show aggressive initial responses with overshoots of 57.87\% and 46.47\% respectively, but successfully converge with minimal steady-state error. PPO and TD-MPC2 adopt conservative strategies, limiting maximum control efforts to 8.58 and 12.48 respectively.

Table~\ref{tab:oc_results} quantifies the performance differences. TD3 achieves optimal tracking (ISE = 3.74, ITAE = 2.66), whilst PPO prioritises control economy (IACE = 2.02) with superior stability margins (GM = 7.55 dB, DM = 0.37s). Notably, the DRL algorithms were not trained for these extreme conditions, yet demonstrate impressive adaptability in handling simultaneous saturations, unmodelled dynamics, and disturbances, validating their implicit robustness acquired through stochastic training environments.

\begin{table}[t!]
\caption{Controller Simulation results, under operational challenges for the AUV System}
\label{tab:oc_results}
\centering
\setlength{\tabcolsep}{8pt}
\renewcommand{\arraystretch}{1.1}
\begin{tabular}{lcccc}
\hline
\textbf{Metric} & \textbf{DDPG} & \textbf{TD3} & \textbf{PPO} & \textbf{TD-MPC2} \\
\hline
$t_r$ (s) & 1.32 & 1.34 & 2.14 & 2.07 \\
$M_p$ (\%) & 57.87 & 46.47 & 23.28 & 20.00 \\
$t_s$ (s) & 17.69 & 23.12 & 24.98 & 24.58 \\
$e_{\mathrm{ss}}$ & 0.04 & 0.05 & 0.10 & 0.05 \\
ISE & 4.82 & 3.74 & 4.57 & 3.73 \\
ITAE & 3.02 & 2.66 & 5.18 & 2.95 \\
IACE & 2.44 & 3.51 & 2.02 & 2.94 \\
IACER & 139.84 & 280.14 & 174.32 & 333.24 \\
$u_{\max}$ & 20.00 & 19.57 & 8.58 & 12.48 \\
GM & 2.80 & 3.15 & 7.55 & 3.28 \\
DM & 0.16 & 0.11 & 0.37 & 0.22 \\
\hline
\end{tabular}
\end{table}

\subsubsection{Performance Tradeoffs and Practical Implications}

The AUV benchmark reveals fundamental engineering tradeoffs between transient performance, control authority requirements, and robustness to operational constraints. Under nominal conditions, TD3 achieves exceptional settling times and error minimisation at the cost of substantial control effort. However, the operational challenges assessment reveals a critical distinction: whilst classical LQI control provides conservative nominal performance, it exhibits catastrophic failure under combined saturations and model uncertainty (Rohrs test), highlighting fundamental limitations of linear control for constrained nonlinear systems.

DRL algorithms demonstrate resilience across all operating conditions. DDPG provides an optimal compromise between aggressive performance and robustness, maintaining stable operation even under severe operational constraints. PPO's conservative strategy, initially appearing suboptimal in nominal conditions, proves advantageous under challenging scenarios with superior stability margins (GM = 7.55 dB). TD-MPC2's model-based approach achieves balanced performance, though with increased control activity under perturbations.

The stark contrast between nominal and constrained operation underscores a critical insight for maritime applications: robustness to operational constraints outweighs nominal performance metrics. The DRL algorithms' ability to implicitly handle saturations, unmodelled dynamics, and disturbances without explicit anti-windup or robust control modifications represents a paradigm shift from classical approaches. This capability proves essential for autonomous marine systems operating in unpredictable oceanic environments where actuator limits, model uncertainties, and environmental disturbances are unavoidable rather than exceptional conditions.

\subsection{Crazyflie Quadrotor}
\begin{figure}[!h]
\centering
\includegraphics[width=0.6\linewidth]{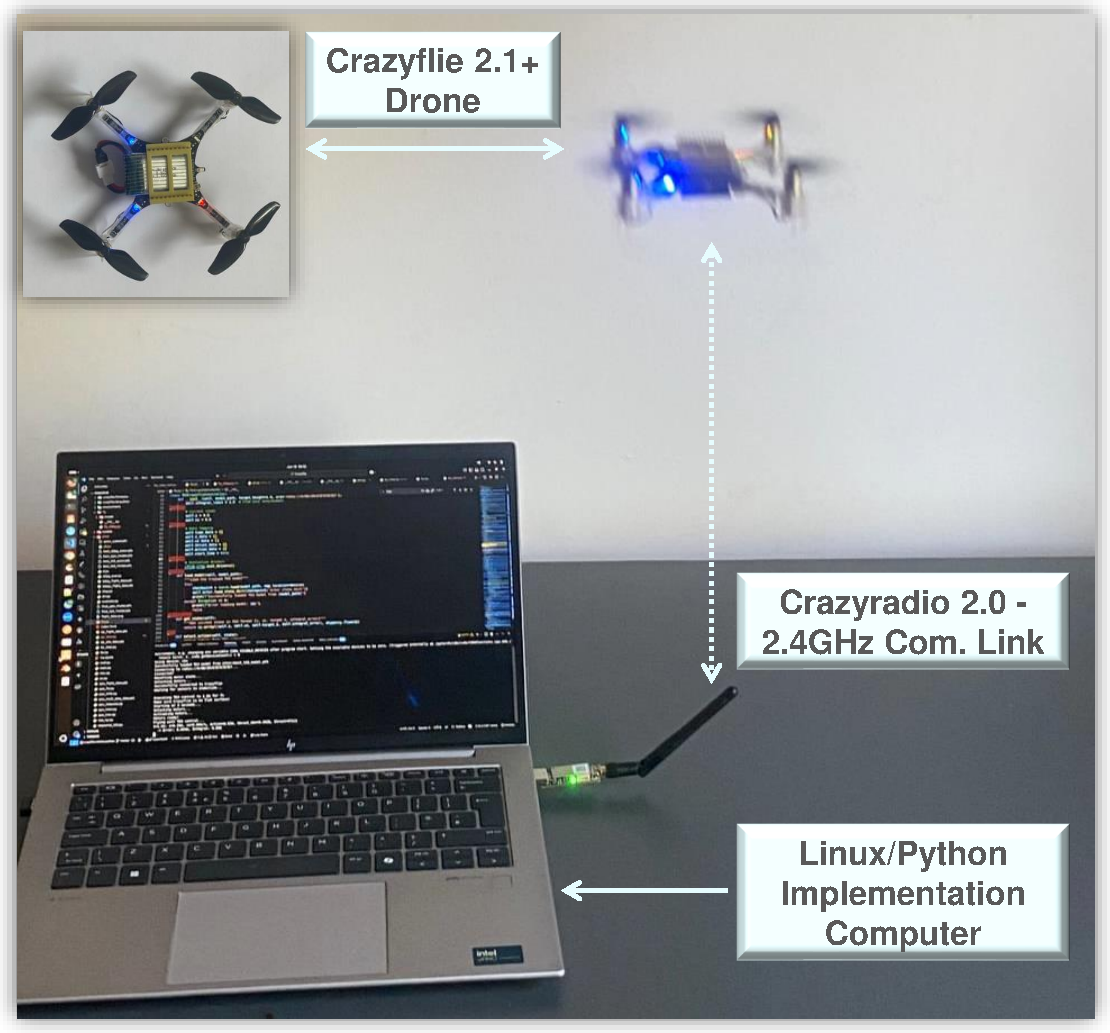}
\caption{Experimental setup for real-world validation of trained DRL controllers on the Crazyflie platform}
\label{fig:crazyflie_setup}
\end{figure}

The final benchmark test evaluates the real-time implementability of the controllers for the altitude control problem of a quadrotor drone, an aspect not considered in such detail in other papers. The Crazyflie 2.1+ nano-quadrotor \cite{crazyflie21datasheet} represents a challenging benchmark, featuring inherent instability, fast dynamics, and strict real-time constraints that mirror industrial UAV control requirements. Figure~\ref{fig:crazyflie_setup} presents the experimental configuration enabling direct validation of simulation-trained controllers on physical hardware.

The complete Crazyflie dynamics comprise a 12-state nonlinear model capturing full 6-DOF motion with complex aerodynamic coupling \cite{nguyen2023crazyflie}. However, to evaluate the generalisability and adaptability of the algorithms, we employ a reduced-order linear model that retains key control characteristics while eliminating nonlinearities, to assess whether the controllers can demonstrate simulation-to-real transfer capability. 

The vertical motion of this vehicle can be modelled using the Newton's second law: $m\ddot{z} = T - mg$, where $z$ represents altitude, $m$ is the drone mass, $g$ is the gravitational acceleration, and $T$ denotes the total thrust of propellers. The goal is to control altitude $z$ by applying thrust $T$, with gravity $g$ treated as a constant disturbance compensated by feedforward thrust during takeoff manoeuvre. Therefore, the vertical dynamics of the drone can be represented in state-space form as:

\begin{equation}
\large
\begin{aligned}
\begin{bmatrix}
\dot{z} \\
\ddot{z}
\end{bmatrix} &= \begin{bmatrix}
0 & 1 \\
0 & 0
\end{bmatrix} \begin{bmatrix}
z \\
\dot{z}
\end{bmatrix} + \begin{bmatrix}
0 \\
\frac{1}{m}
\end{bmatrix} T + \begin{bmatrix}
0 \\
-g
\end{bmatrix},
\\
y &= z.
\end{aligned}
\end{equation}

where the state vector comprises the altitude $z$ and the vertical velocity $\dot{z}$, with the mass $m = 0.027$ kg.

\subsubsection{Performance Validation}
The challenging multi-step reference sequence $0 \to 0.5\text{m} \to 1.0\text{m} \to 0.5\text{m}$ systematically evaluates tracking consistency across operating points whilst revealing the differences between idealised simulation and physical implementation. This comprehensive evaluation protocol provides unique insights into the practical viability of simulation-trained DRL controllers for safety-critical aerial systems.

\begin{figure}[!h]
\centering
\includegraphics[width=1.0\textwidth]{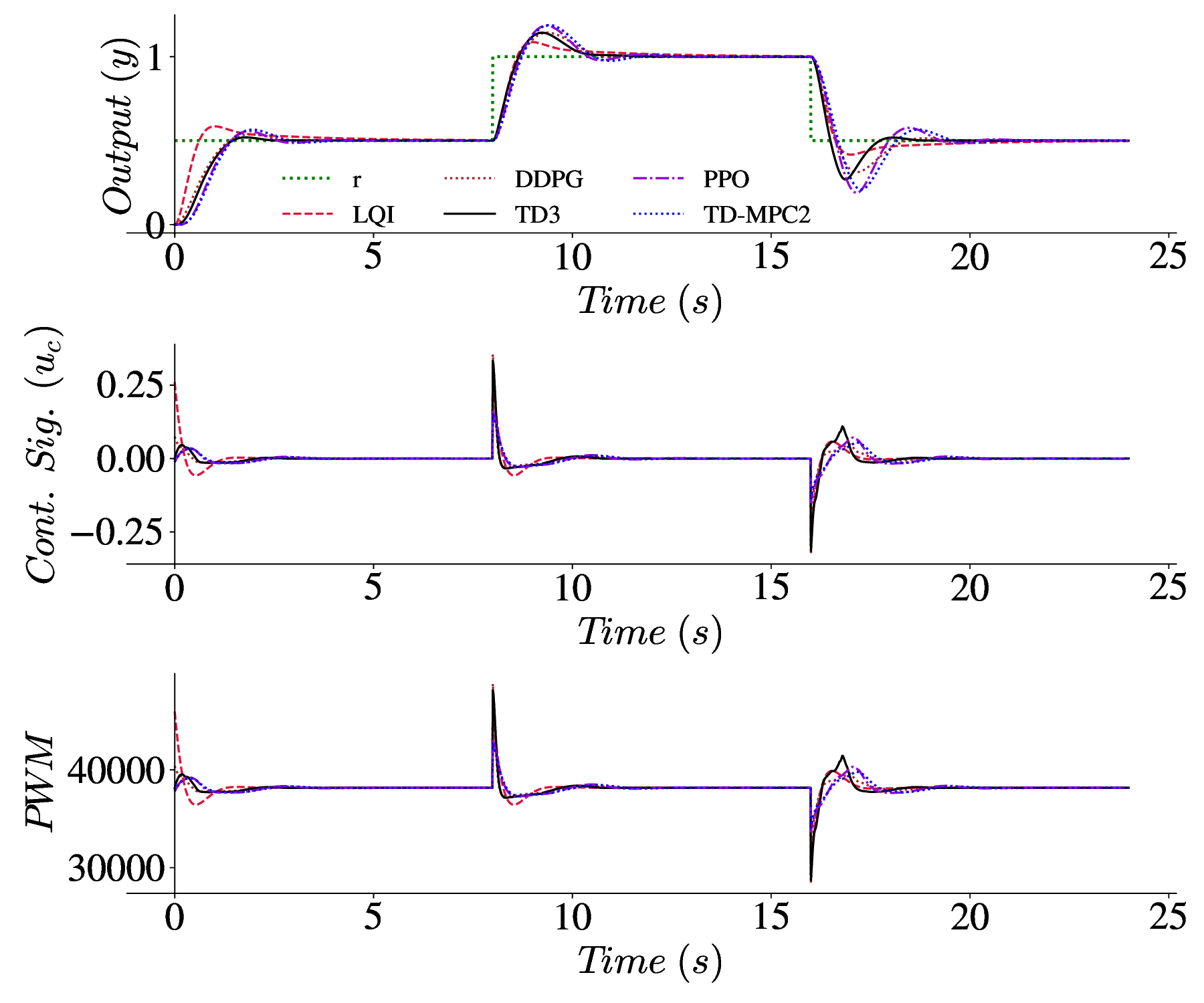}
\caption{Multi-step reference tracking comparison (Simulation)}
\label{fig:crazyflie_sim}
\end{figure}

\begin{figure}[!h]
\centering
\includegraphics[width=1.0\textwidth]{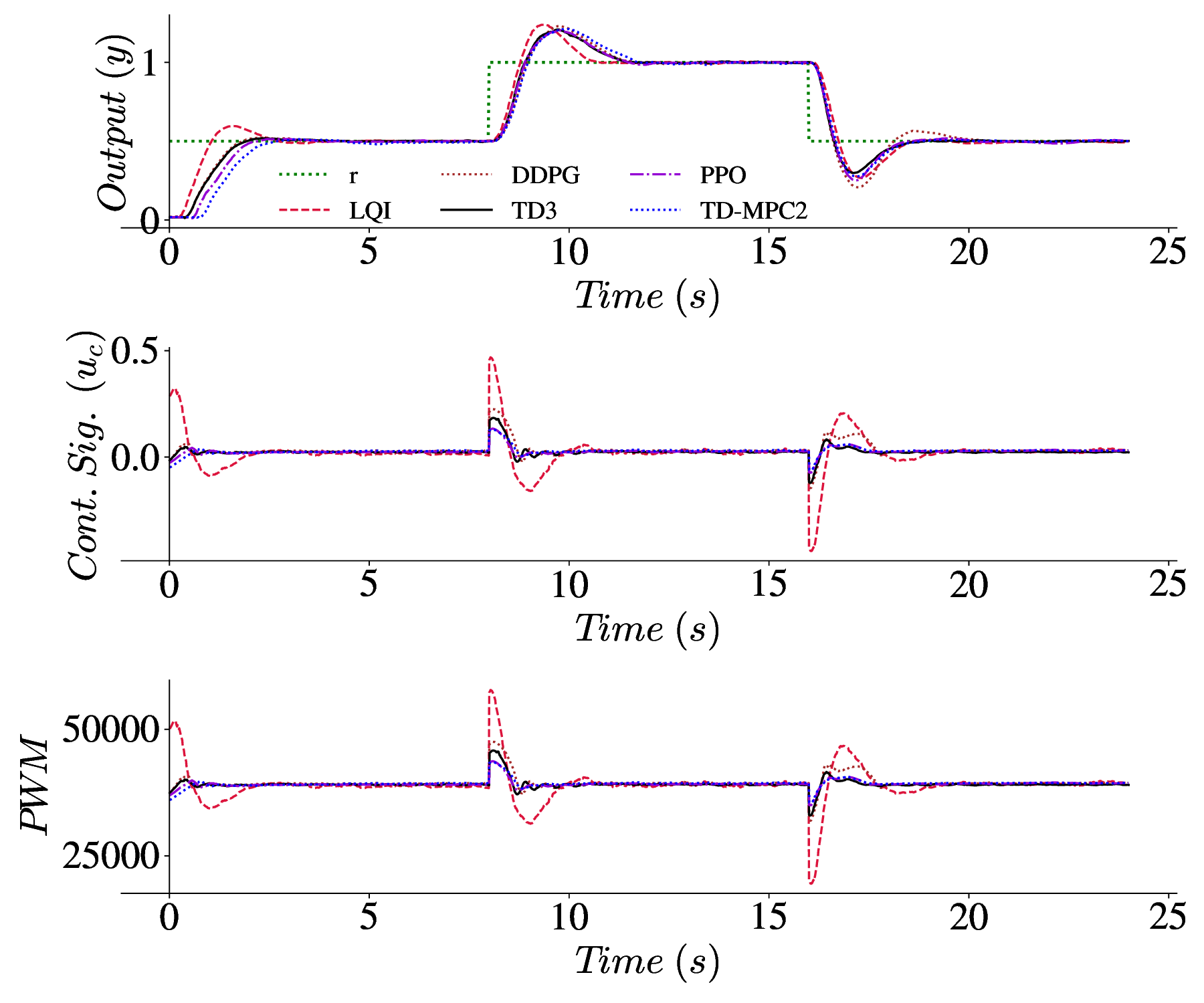}
\caption{Multi-step reference tracking comparison (Experimental Validation)}
\label{fig:crazyflie_exp}
\end{figure}

Figure~\ref{fig:crazyflie_sim}, \ref{fig:crazyflie_exp} presents a comparison validating simulation fidelity whilst highlighting the inevitable complexities of real-world deployment. The simulation results Figure~\ref{fig:crazyflie_sim} demonstrate clean tracking performance across all algorithms, with quantitative metrics presented in Table~\ref{tab:crazyflie_sim_results}. TD3 achieves the most aggressive transient response with rapid settling (2.6s) despite moderate overshoot (37.5\%), whilst LQI provides the most conservative approach with longer settling times (4.9s) but acceptable overshoot characteristics (17.0\%). PPO exhibits the highest overshoot (49.9\%) among all controllers, reflecting its stochastic policy exploration, whilst TD-MPC2 demonstrates balanced performance with overshoot (42.6\%) and competitive settling behaviour.

\begin{table}[t!]
\caption{Performance Results for Crazyflie Multi-Step Reference Tracking (Simulation)}
\label{tab:crazyflie_sim_results}
\centering
\small
\setlength{\tabcolsep}{6pt}
\renewcommand{\arraystretch}{1.0}
\begin{tabular}{lccccc}
\hline
\textbf{Metric} & \textbf{LQI} & \textbf{DDPG} & \textbf{TD3} & \textbf{PPO} & \textbf{TDMPC2}\\
\hline
$t_r$ (s) & 0.23 & 0.20 & 0.25 & 0.30 & 0.21 \\
$M_p$ (\%) & 17.00 & 33.20 & 37.50 & 49.90 & 42.60 \\
$t_s$ (s) & 4.90 & 2.90 & 2.60 & 4.10 & 3.30 \\
$e_{\text{ss}}$ & 0.00 & 0.00 & 0.00 & 0.00 & 0.00 \\
ISE & 0.83 & 1.18 & 1.20 & 2.32 & 1.52 \\
ITAE & 37.11 & 39.28 & 33.93 & 60.31 & 47.64 \\
IACE & 0.79 & 0.66 & 0.81 & 0.83 & 0.71 \\
IACER & 6.92 & 6.90 & 7.37 & 3.98 & 4.37 \\
$u_{\max}$ & 0.30 & 0.40 & 0.30 & 0.10 & 0.20 \\
\hline
\end{tabular}
\end{table}

The control effort analysis reveals distinct characteristics among algorithms. TD3 and LQI demonstrate similar maximum control demands (0.30), whilst PPO exhibits remarkably conservative control usage (0.10) at the cost of increased tracking errors. The integral control effort rates (IACER) indicate that LQI maintains the smoothest control signals (6.92), with TDMPC2 providing competitive smoothness (4.37) through its trajectory optimisation approach.

\subsubsection{Real-World Implementation Results}
The experimental validation on actual Crazyflie hardware (Table~\ref{tab:crazyflie_real_results}) reveals the inevitable but manageable performance degradation when transitioning from simulation to physical implementation. Most significantly, all algorithms demonstrate increased rise times compared to simulation, with TDMPC2 showing the most pronounced degradation (0.21s to 0.80s), suggesting sensitivity to model uncertainties inherent in real-world dynamics.

\begin{table}[!t]
\caption{Performance Results for Crazyflie Multi-Step Reference Tracking (Real Hardware)}
\label{tab:crazyflie_real_results}
\centering
\small
\setlength{\tabcolsep}{6pt}
\renewcommand{\arraystretch}{1.0}
\begin{tabular}{lccccc}
\hline
\textbf{Metric} & \textbf{LQI} & \textbf{DDPG} & \textbf{TD3} & \textbf{PPO} & \textbf{TDMPC2}\\
\hline
$t_r$ (s) & 0.30 & 0.64 & 0.62 & 0.60 & 0.80 \\
$M_p$ (\%) & 30.00 & 28.70 & 21.70 & 24.40 & 22.90 \\
$t_s$ (s) & 3.60 & 3.50 & 3.00 & 4.30 & 4.00 \\
$e_{\text{ss}}$ & 0.00 & 0.00 & 0.00 & 0.00 & 0.00 \\
ISE & 1.60 & 2.06 & 1.80 & 2.00 & 2.30 \\
ITAE & 53.40 & 63.10 & 50.10 & 55.20 & 58.70\\
IACE & 4.40 & 3.40 & 2.60 & 2.70 & 2.90 \\
IACER & 18.0 0& 8.00 & 6.40 & 4.3 & 4.00 \\
$u_{\max}$ & 0.50 & 0.20 & 0.20 & 0.10 & 0.10 \\
\hline
\end{tabular}
\end{table}

Remarkably, the overshoot characteristics demonstrate convergence in real-world implementation, with all algorithms exhibiting similar peak overshoots (21.7\%-30.0\%) compared to the wider variation observed in simulation (17.0\%-49.9\%). This convergence suggests that physical constraints, including actuator limitations, sensor noise, and aerodynamic effects, naturally limit aggressive control behaviour, providing an inherent robustness mechanism absent in idealised simulation environments.

The control effort analysis reveals a critical insight: real-world implementation demands significantly higher control activity (IACE increases by 3-5 times across all algorithms) and control rates (IACER increases by 2-3 times). This phenomenon reflects the continuous correction required to compensate for sensor noise, aerodynamic disturbances, and model uncertainties. TD3 maintains the best settling performance (3.0s) whilst requiring moderate control effort, demonstrating superior practical robustness among DRL approaches.

\subsubsection{Sim-to-Real Transfer Analysis}
The quantitative comparison between simulation and experimental results provides valuable insights into sim-to-real transfer effectiveness. The most significant degradation occurs in integral time-weighted error metrics (ITAE), increasing by 35-60\% across all algorithms, indicating that whilst fundamental tracking capability is preserved, transient response quality inevitably suffers under real-world conditions.

Critically, all algorithms maintain zero steady-state error in both domains, validating the integral action component of the state representation and confirming the robustness of the fundamental control architecture. The successful completion of the multi-step reference sequence across all controllers demonstrates that simulation-trained policies possess sufficient robustness margins to handle unmodelled dynamics, sensor noise, and environmental disturbances characteristic of practical UAV operations.

The achievement of stable, repeatable flight performance across fundamentally different learning paradigms, from classical optimal control to cutting-edge model-based reinforcement learning, establishes a compelling validation of DRL viability for safety-critical aerial applications. This comprehensive evaluation provides quantitative evidence that current DRL techniques have matured sufficiently for practical deployment in demanding real-world scenarios, whilst highlighting the importance of robust training methodologies that account for inevitable implementation challenges.

\section{Conclusion}
This paper presented several months of investigation into the achievable performance of some widely discussed deep reinforcement learning controllers, a topic still lacking structured evaluation in the existing literature. It considered key practical control challenges such as noise, disturbances, model uncertainty, actuator rate and amplitude saturation, time delay, and real-time applicability. These issues were examined through detailed black-box analyses using tailored evaluation metrics. The study evaluated these controllers on a range of well-known benchmarks: a non-minimum-phase process control example, the challenging ACC two-mass-spring problem, nonlinear REMUS AUV yaw control, and real-time deployment on a quadrotor's altitude controller. The work provides a clear insight into the practical performance and robustness of DRL approaches, as outlined below:

\begin{itemize}[noitemsep, topsep=0pt]
\item DRL algorithms, without requiring a system model, are capable of learning control policies that offer performance comparable with classical model-based controllers.
\item They are able to manage real-world challenges such as delay, non-minimum-phase behaviour, actuator constraints, and uncertainties resembling those found in practical systems.
\item These methods respond well to actuator saturation and windup phenomenon, even when such effects are not considered during training.
\item They are implementable in real time, even when the sampling rate differs from that used during training.
\item DRL shows reasonable capability in transferring from simulation to physical implementation for learning on a non-precise model.
\item Although hyperparameter tuning is time-intensive, these algorithms show promise in handling higher-level decision-making and control tasks through appropriate integration into their reward design.
\item Despite these successes, the achievable performance is not significantly better than that of a simple controller such as gain state feedback; however, it demonstrates promise for the aforementioned higher-level control problems.
\end{itemize}

TD3 demonstrated strong performance in time-domain response, whereas PPO showed more consistent behaviour with good stability properties. However, no single controller excelled in every aspect, highlighting the importance of selecting algorithms based on the specific demands of each application. For example, TD-MPC2 demonstrates better performance under actuator limitations. Consequently, controller design becomes closely tied to the application and tuning process. With sufficient tuning effort, comparable performance can be achieved for DDPG, TD3, and PPO. Future work, as already being explored in the literature, involves establishing formal guarantees for these algorithms; nevertheless, any such advancement must be evaluated through comprehensive testing, as performed in this study, to determine real-world applicability. Furthermore, future algorithms should seek significant improvements in architecture and control performance, or adopt more intelligent reward function design, since minor improvements in accumulated reward during training may not translate to meaningful benefits in practice.

\bibliographystyle{elsarticle-num}
\bibliography{main}
\end{document}